# Carrier Heating and Negative Photoconductivity in Graphene

J. N. Heyman[1], J. D. Stein[1], Z. S. Kaminski[1], A. R. Banman[1], A. M. Massari[2], J.T. Robinson[3]

1. Macalester College, St. Paul, MN 55105
2. School of Chemistry, Univ. Minnesota, Minneapolis MN 55455
3. Naval Research Laboratory, Washington DC 20375

## ABSTRACT

We investigated negative photoconductivity in graphene using ultrafast terahertz techniques. Infrared transmission was used to determine the Fermi energy, carrier density and mobility of *p*-type CVD graphene samples. Time-resolved terahertz photoconductivity measurements using a tunable mid-infrared pump probed these samples at photon energies between 0.35eV – 1.55eV, approximately one-half to three times the Fermi energy of the samples. Although interband optical transitions in graphene are blocked for pump photon energies less than twice the Fermi energy, we observe negative photoconductivity at all pump photon energies investigated, indicating that interband excitation is not required to observe this effect. Our results are consistent with a thermalized free carrier population that cools by electron-phonon scattering, but inconsistent with models of negative photoconductivity based on population inversion.

## I. INTRODUCTION

Graphene is a single atomic sheet of covalent *sp*$^2$ bonded carbon atoms that forms a zero-gap semiconductor with linear bands. Graphene has extraordinary mechanical, electronic and optical properties. In opto-electronics the material may find applications as a modulator of terahertz or infrared radiation due to strong free carrier absorption that is gate tunable, or as a saturable absorber due to strong nonlinear optical effects. Graphene may also be exploited as a transparent conducting layer due to its high conductivity and weak optical absorption due to interband transitions. Recent research has also explored graphene as a source of THz or infrared radiation, including as a possible gain medium in a THz laser[1].

A variety of time-resolved probes have been used to study hot carrier dynamics in graphene. In these studies, the material is excited with an optical or infrared pump pulse and the system is probed with photoelectron[2-5] or optical spectroscopy at visible[6-10], infrared[11, 12] or THz[13-23] frequencies. The majority of these studies are consistent with a single model: Carrier-carrier and carrier-phonon intraband scattering produces distinct Fermi distributions of electrons and holes ~50fs following photoexcitation. Interband scattering merges these distributions, producing a thermal distribution of carriers in the valance and conduction band with a common chemical potential within ~300fs. Carrier-phonon scattering causes the hot electrons and holes to equilibrate with the optical phonons within ~1ps. The ensemble of hot carriers and phonons cools to equilibrium as the optical phonons decay over longer timescales.

Recent research has probed whether optical excitation can create a population inversion between the valance and conduction bands in graphene, as is commonly seen in direct-gap semiconductors such as GaAs. In zero-gap graphene, this could result in gain at THz frequencies, which could be the basis of an optically pumped THz laser. According to the model above, this is only possible within ~300fs following excitation. At later times the carrier distribution is thermal and a population inversion is not possible. Ultrafast photoelectron spectroscopy by Gierz[4] has tracked electron populations near the Dirac point following mid-infrared or near-infrared photoexcitation. Near-infrared excitation ($hf > 2\varepsilon_F$) excites interband transitions. Rapid intraband scattering thermalizes the electron and hole populations within <30fs, but they find that the chemical potential in the valance and conduction bands are distinct, with a population inversion which persists for ~130fs. At later times the carriers in both bands are described by a Fermi distribution with a single carrier temperature and chemical potential. On longer timescales (0.1-5ps) the carriers cool and recombine. As discussed below, excitation with mid-infrared photons ($hf < 2\varepsilon_F$) drives only intraband transitions in their *p*-type samples. They find that ~30fs after mid-infrared photoexcitation the free carrier distribution is described by a Fermi function with a single well-defined carrier temperature and chemical potential due to rapid electron-electron and electron-phonon scattering. Photoelectron spectroscopy of graphene with NIR excitation by Johannsen[2] did not resolve dynamics on timescale <100fs, and supports the carrier cooling picture on longer timescales. Measurements with NIR and with VIS excitation by Gilbertson[3] identified distinct electron and hole chemical potentials which persist for ~500fs, after which the electron and hole distribution functions merge.

Time resolved optical studies of graphene are largely consistent with these photoemission results. Li, et. al.,[12] performed time-resolved two-color near-infrared reflectivity measurements following near-infrared excitation. They report a broadband inverted carrier population and optical gain that persist for ~200fs after excitation. At later times the electron and hole distribution functions merge. Breusig[8] studied single exfoliated graphene layers and observed an initially non-thermal carrier distribution which evolves to a thermal distribution in ~200fs, but did not report a population inversion. Earlier two-color time-resolved transmission and reflection measurements by Newsom[9], and by Sun[11] are consistent with these results, reporting carrier thermalization within ~100fs of excitation followed by rapid carrier cooling by optical phonon emission (0.2-0.5ps) and slower cooling by acoustic phonon scattering and decay of hot optical phonons into acoustic phonons (1-4ps). Degenerate pump-probe measurements by Dawlenty[10] also report carrier thermalization within a few hundred femtoseconds of optical excitation followed by carrier cooling.

Time-resolved THz spectroscopy has also been applied to graphene. At THz frequencies the conductivity is typically dominated by the Drude conductivity of free carriers (i.e. intraband transistions) rather than interband transistions, which are suppressed due to Pauli blocking at low frequencies in extrinsic graphene ($hf < 2\varepsilon_F$). In time-resolved studies the samples are excited with a pump pulse at optical, infrared or THz pump frequencies, and the resulting change in transmission of a delayed THz probe pulse is used to measure the change in conductivity. The temporal resolution of THz probe spectroscopy is ~0.2ps.

George, et. al. [13] used optical pump, THz probe measurements to study epitaxial graphene on Si-C. He observed a rapid decrease in transmission (increase in conductivity) within 500fs of photoexcitation, followed by a recovery to equilibrium over the next 1-5ps. He was able to fit the observed dynamics with model that assumed that a hot, thermalized carrier distribution forms within ~150fs of photoexcitation. The dynamics at later times reflect the cooling of the electrons by intraband optical phonon scattering over the timescale 0.15-1ps and by other interband recombination processes at still longer times. Choi[14] presented a similar study of epitaxial graphene that was also consistent with a thermal model. Wang[15] reported similar results from epitaxial graphene and CVD graphene grown on nickel. They were able to model the cooling of the photoexcited electron-hole plasma using a set of coupled rate equations to fit their data, including pump-power dependence.

Boubanga-Tombet[16, 17] obtained distinctly different results in a 1550nm NIR pump, near-field THz probe measurement on graphene. They report that THz pulses that probe the graphene sample 2ps and 3.5ps after photoexcitation are amplified by up to ~50%, and they ascribe the gain to stimulated emission due to an inverted population. They report a threshold for gain of ~$10^7$W/cm$^2$, which is of the same order as reported for population inversion in the photoelectron and optical studies above. However, the population inversion responsible for gain would need to persist for at least ~4ps to explain their results, an order of magnitude longer than observed in other experiments.

More recently, negative differential conductivity has been observed in optical-pump, THz probe measurements on graphene samples grown by CVD on copper. Here, optical excitation by a pump pulse *increases* the transmission of a delayed terahertz (THz) probe pulse, indicating a transient *decrease* in conductivity. Docherty[24] observed an increase in the THz transmission in graphene following photoexcitation that decayed on a few-ps timescale which they interpreted as gain due to stimulated emission (a negative interband conductivity). They also observed a strong sensitivity of this effect to the sample environment. Jnawali[18] also observed a transient decrease in THz conductivity of CVD graphene following photoexcitation in similar experiments. They showed that in highly doped, relatively high mobility graphene samples, the increase in electron-phonon scattering in hot photoexcited graphene samples dominates over the effect of an increase in carrier density resulting in an overall decrease in the *intraband* conductivity.

Tielrooij, et. al. [20], performed optical-pump, THz probe measurements on CVD graphene using a tunable pump pulse which was varied between 4.65 eV and 0.16eV. They observed negative differential conductivity (transmission increase following photoexcitation) that they also ascribed to a transient increase in scattering rate of photoexcited carriers. In addition, they observed that the peak conductivity change scaled with linearly with pump-photon energy and absorbed photon flux. This indicates that the scattering process that drives the initial carrier thermalization redistributes the absorbed energy between the carriers rather than removing energy from the system, indicating that initial thermalization is dominated by carrier-carrier scattering rather than by electron-phonon scattering. Optical pump-THz probe measurements in gated graphene samples by Shi, et. al.,[22] show that photoconductivity can be tuned from positive to negative with increasing carrier concentration find this to be consistent with the carrier cooling model.

Hwang[21] reported THz pump, THz probe measurements that showed a strong pump-induced conductivity decrease (transmission increase) in CVD graphene samples. The THz pump can only drive intraband excitation, and should heat the carrier distribution without producing a population inversion, so this showed that population inversion is not required to observe negative differential conductivity in these experiments.

Recent reports of photoluminescence in graphene[25] also support a thermal carrier distribution. The authors observe broad band black-body emission from the photocarriers after photoexcitation. The results indicate that carriers in graphene are well thermalized among themselves within ~100fs of excitation.

In this work we investigate transient photoconductivity in graphene following an infrared pump pulse. We are able to tune the pump pulse to photon energies where interband excitation is not possible, due to Pauli blocking. If negative photoconductivity in our samples were due to interband gain, it should vanish under these conditions. However, we see negative photoconductivity at all pump photon energies investigated, and find that the photo-induced conductivity change depends only on the amount of pump energy absorbed. We are able to fully describe our results with a thermal model, in which free electrons and holes have a common temperature and chemical potential throughout the measurement, and the dynamics are described by the cooling of the free carriers by electron-phonon scattering. While our results are similar to those of Tielrooij, et. al. and yield broadly similar results, our work investigates carrier dynamics in samples which have been characterized by infrared spectroscopy to determine carrier density, scattering rate and Fermi energy.

## II.     EXPERIMENT

Single layer graphene samples were grown on copper foil by chemical vapor deposition (CVD) following Li *et. al.* [26]. After growth the copper foil was dissolved and the graphene films were transferred to 1cm$^2$ sapphire substrates for further measurements. Electrical resistivity and Hall Effect measurements showed the samples to be *p*-type with carrier concentrations in the range $p = 2 \cdot 10^{12}$cm$^{-2}$ - $2 \cdot 10^{13}$cm$^{-2}$, and DC mobilities $\mu =$ 1000-2000 cm$^2$/Vs. Multilayer CVD-graphene samples grown on nickel obtained from Graphene Supermarket were also transferred to sapphire substrates. Atomic Force Microscopy on these indicated a thickness of 15-20 layers.

Infrared transmission measurements were performed using a Thermo-Nicolet IS-50 FTIR with near-, mid-, and far-infrared optics. Far-infrared transmission measurements used a 4.2 K Si bolometer (IR Labs) as a detector. Time resolved terahertz transmission measurements were performed using a THz system integrated into the University of Minnesota Multi-User Laser Experiment (MULE), which is based on an amplified Ti-Sapphire laser (1KHz repetition rate, 3mJ/pulse, 120fs pulses at $\lambda$ = 800nm). THz probe pulses were generated by optical rectification of the 800nm pulses in a 1mm ZnTe crystal and focused onto the sample. The sample was excited with pump pulses from an optical parametric amplifier driven by the Ti:S laser. The pump pulses were tunable, with $\hbar\omega = 0.3 - 1.1$ eV in this study. Transmitted radiation was focused onto a 1mm ZnTe electro-optic detector probed with a delayed component of the 800nm pulse. The pump was modulated with a chopper and the change in THz transmission was measured with a lock-in amplifier. The THz beam path was purged to suppress absorption by water vapor. Additional TRTS measurements were performed at Macalester College in a THz system based on a FemtoLasers XL500 chirped pulse Ti:S oscillator (5MHz repetition rate, 0.5$\mu$J/pulse, 50fs). The pump pulse energy system in this system is fixed at $hf$ = 1.55eV and we used a photoconductive switch as a THz emitter. Infrared and THz measurements interrogated 3mm diameter regions on the sample.

III. RESULTS

Infrared transmission spectra of our samples (Fig. 1) show free-carrier absorption and interband absorption typical of graphene[27]. Model fits to the data yield the hole concentration $p = 1.1 \cdot 10^{13}$ cm$^{-2}$ and carrier scattering time $\tau$ = 77fs for single-layer sample JR1203. For the multi-layer sample GS0920 we found $p = 3 \cdot 10^{13}$ cm$^{-2}$ and carrier scattering time $\tau$ = 30fs. We modeled the infrared transmission due to the graphene as follows:

$$T = 1 - A_{INTER} f(-\hbar\omega/2)(1 - f(\hbar\omega/2)) - A_{DRUDE} \left( \frac{\tau}{1 + \omega^2 \tau^2} \right), \qquad (1)$$

where

$$A_{INTER} = \frac{2\pi\alpha}{n_s + 1} \quad ; \quad A_{Drude} = \left( \frac{2\mu_0 c}{n_s + 1} \right) \frac{D}{\pi} \frac{\tau}{1 + \omega^2 \tau^2} s . \qquad (2)$$

Here, $\alpha$ is the fine structure constant, $n_s$ is the substrate index of refraction, $\mu_0$ is the vacuum permeability, $1/\tau$ is the carrier scattering rate and $D = e^2 v_F \sqrt{\pi p}$ is the Drude strength. The expression for the Drude absorption is valid to first order in the transmission change, and we found no important differences by fitting to an exact expression. The scaling factor $s$ accounts for the reduction in intraband absorption due to the finite interband absorption, given that these satisfy a sum rule.[28] We observe a cutoff in the interband absorption for photon energies $hf < 2\varepsilon_F$ due to initial state occupancy because the Fermi energy $\varepsilon_F \gg k_B T$ in our CVD graphene samples. Setting $s = 0.8$ makes the Drude absorption strength consistent with the cutoff in interband absorption

Figure 2 shows the pump-induced change in transmission in two graphene samples as a function of delay between the arrival of the pump pulse ($hf = 1.55$eV) and the THz probe pulse. The sign of transmission change is sample dependent. The single layer sample JR1104 grown on copper shows an *increase* in transmission following pump excitation (negative photoconductivity), while multi-layer sample GS0920 shows a *decrease* in THz transmission (positive photoconductivity). The model fits to the data are discussed below.

We measured the pump-induced transmission change in the single-layer graphene sample JR1203 as a function of delay for pump photon energies between 0.35eV and 1.55eV (Fig. 3) The pump fluence incident on the samples was ~0.2 mJ/cm² for all pump energies except 1.55eV and 0.35eV, where it was approximately one order of magnitude lower. We observe negative photoconductivity at all pump wavelengths investigated. Importantly, measurements obtained using pump-photon energies above and below the cut-off for interband absorption show no qualitative differences. We have also graphed the pump-induced transmission versus pulse energy (Fig. 4) for pump photon energies 1.55eV ($\hbar\omega \gg 2\varepsilon_F$) and 0.52eV ($\hbar\omega \ll 2\varepsilon_F$). When we scale the pump power by the linear absorbance we find that for absorbed pump energies below ~30nJ/cm², the photoconductive sensitivity is approximately equal at the two pump wavelengths. At higher powers we observe saturation at pump energy 1.55eV, presumably because of saturation of interband absorbance.

IV. DISCUSSION

We have adapted a model of transient photoconductivity in graphene by Rana, *et. al.*, [29] to describe our measurements. The model assumes that photoexcited electrons and holes thermalize within ~100fs of photoexcitation. At later times the electron and hole distribution functions are described by a single carrier temperature and chemical potential. The hot carriers cool by optical phonon scattering, and the model tracks the total kinetic energy of the free carriers $Q(t)$, carrier temperature, electron and hole carrier densities, phonon populations, and carrier scattering rates to determine the conductivity versus time.

The background doping and the initial energy provided by the laser pump pulse $Q(0)$ determine the initial carrier temperature and the electron and hole distribution functions. We assume that $\Gamma$-point and $K$-point optical phonons dominate intra-valley and inter-valley scattering respectively. For an electron at energy $\varepsilon$ the scattering rates for $\Gamma$ optical phonon absorption, $G_{\Gamma e}(\varepsilon)$, and emission, $R_{\Gamma e}(\varepsilon)$, are given by:

$$G_{\Gamma e}(\varepsilon) = 2A \left| \frac{\varepsilon + \hbar\omega_\Gamma}{\hbar\omega_\Gamma} \right| (1 - f_e(\varepsilon + \hbar\omega_\Gamma)) N_\Gamma \quad (3)$$

$$R_{\Gamma e}(\varepsilon) = 2A\left|\frac{\varepsilon - \hbar\omega_\Gamma}{\hbar\omega_\Gamma}\right|(1 - f_e(\varepsilon - \hbar\omega_\Gamma))(1 + N_\Gamma) ,$$

where $A = \frac{9}{4}\left(\frac{\partial^2 t}{\partial b^2}\right)\frac{\hbar\omega_\Gamma}{\pi\rho\hbar^4 v_F^4}$ , $\omega_\Gamma$ is the $\Gamma$-optical phonon frequency, $\rho$ is the density of graphene and $\frac{\partial^2 t}{\partial b^2}$ is defined by Rana[29]. A similar expression describes carrier scattering involving intervalley $K$ phonons, as well as scattering of holes. The rate of energy transfer from the free carriers to the $\Gamma$ phonons is determined by the scattering rates:

$$\frac{dQ_\Gamma}{dt} = \left[\int f_e(\varepsilon)D_e(\varepsilon)\left(R_{\Gamma e}(\varepsilon) - G_{\Gamma e}(\varepsilon)\right)d\varepsilon + \int f_h(\varepsilon)D_h(\varepsilon)\left(R_{\Gamma h}(\varepsilon) - G_{\Gamma h}(\varepsilon)\right)d\varepsilon\right]\hbar\omega_\Gamma , \quad (4)$$

with a similar expression describing energy transfer to $K$ phonons. The rate of change in phonon populations includes phonon emission and absorption due to carrier scattering and decay of optical phonons into acoustic phonons

$$\frac{dN_\Gamma}{dt} = \left(\frac{1}{\hbar\omega_\Gamma M_\Gamma}\frac{dQ_\Gamma}{dt}\right) - \frac{N_\Gamma - N_\Gamma^0}{\tau_\Gamma} , \quad (5)$$

where $\tau_\Gamma$ is the optical phonon lifetime, $N_\Gamma^0$ is the equilibrium phonon population, and $M_\Gamma$ is the effective number of phonon states at the $\Gamma$ point, as described by Rana[29]. A similar expression describes the $K$-phonon population. Finally we calculate the conductivity $\sigma = \sigma_e + \sigma_h$, where

$$\sigma_e = -\tfrac{1}{2}e^2 v_f^2 \int_{-\infty}^{\infty} \frac{\partial f_e}{\partial \varepsilon} g_e(\varepsilon)\tau_e(\varepsilon)d\varepsilon , \quad (6)$$

describes[30] electron conductivity and a similar expression gives $\sigma_h$. Here $1/\tau(\varepsilon) = R_e(\varepsilon) + G_e(\varepsilon) + r_0$, where $r_0$ is the background scattering rate due to charged impurities and other processes and is taken to be constant, and $g_e(\varepsilon)$ is the density of states. We numerically step equations (3) – (6) forward in time to find the conductivity versus time. The amplitude transmission of the THz pulse is determined by the conductivity

$$T_E = \left(1 + \frac{\mu_0 c \sigma}{n_s + 1}\right)^{-1}. \tag{7}$$

Our carrier cooling model accurately describes our transient conductivity data (Fig 2). The equilibrium carrier densities and background scattering rate were taken as adjustable parameters. The best fit values of these parameters agree with the values determined from IR spectroscopy within a factor of two and we ascribe the discrepancy to the simplified treatment of carrier scattering, although the environmental sensitivity of graphene may also contribute to this difference between independent measurements. Importantly, this simple model can yield either positive or negative photoconductivity depending on the choice of parameters. Qualitatively, the lower equilibrium carrier density per layer in sample GS0920 yields *positive* photoconductivity because the conductivity increase due to the increase in free carrier density is larger than the decrease due to the increased scattering rate of hot carriers. The higher equilibrium carrier density in JR1203 yields *negative* photoconductivity because the increase in scattering rate dominates.

A critical prediction of this model is that the transient photoconductivity is a bolometric response that depends only on the amount of energy absorbed by the graphene layer, and not the pump frequency because the THz probe samples a thermalized free carrier density due to its ~0.2ps time-resolution. This is supported by our transient photoconductivity measurements taken with a tunable pump (Fig. 3, Fig. 4). We find the dynamics to be dependent on pump intensity, but independent of pump photon energy, as predicted by our model. In contrast, our data exclude the possibility that the observed negative photoconductivity is due to a photo-induced population inversion leading to gain at THz frequencies. While interband excitation would be required to create a population inversion, we observe negative photoconductivity with pump excitation both above and below the cutoff for interband excitation. We observe no threshold intensity for negative photoconductivity.

## V.    CONCLUSIONS

We have investigated transient photoconductivity in CVD graphene using a tunable pump. Single layer samples show negative photoconductivity at all pump wavelengths investigated, including photon energies above and below the cutoff for interband absorption. The magnitude of the transient conductivity change is determined by the absorbed energy, but is independent of the pump wavelength. We show our results are consistent with a carrier cooling model in which the free electrons and holes are described by a single carrier temperature and chemical potential, and cool by electron-phonon scattering. We do not see evidence for population inversion or gain on the timescales probed by this experiment.

The authors gratefully acknowledge the contribution of Ms. Margaret Molter and Dr. Tim Anglin in designing and building the experimental apparatus, and of Mr. Rhyan Foo Kune in developing the numerical simulations. This material is based upon work supported by the National Science Foundation under Grant No. DMR-1006065, and by CHE-1048560. Research at NRL was supported by Base Programs funded through the Office of Naval Research.

FIGURES

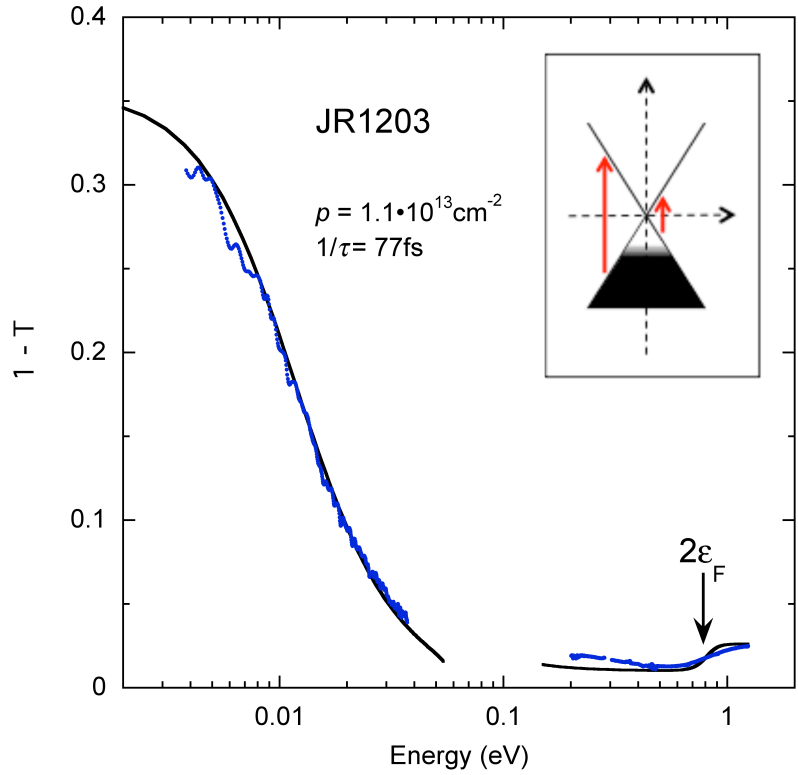

FIG. 1 Infrared transmission spectrum of single layer CVD-graphene on sapphire sample JR1203. The solid line is a model fit. The gap in the data is due to substrate absorption. The inset illustrates interband transitions in our *p*-type samples and shows that interband absorption at $\hbar\omega < 2\varepsilon_F$ is blocked.

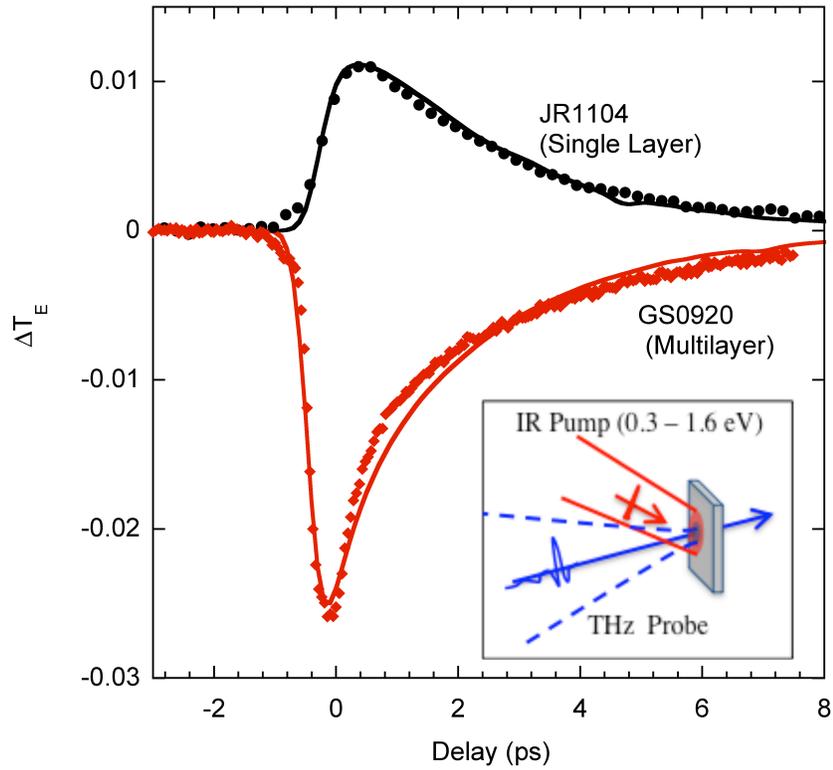

FIG. 2 Time-resolved THz transmission measurements on single layer ($p = 3 \cdot 10^{12}$cm$^{-2}$) and multilayer ($p \sim 2 \cdot 10^{11}$cm$^{-2}$ per layer) CVD graphene samples showing peak amplitude transmission change versus delay between IR pump and THz probe pulses. Solid lines are model fits. The single layer sample shows negative photoconductivity (transmission increase). Inset shows the measurement geometry.

Heyman et. al.,

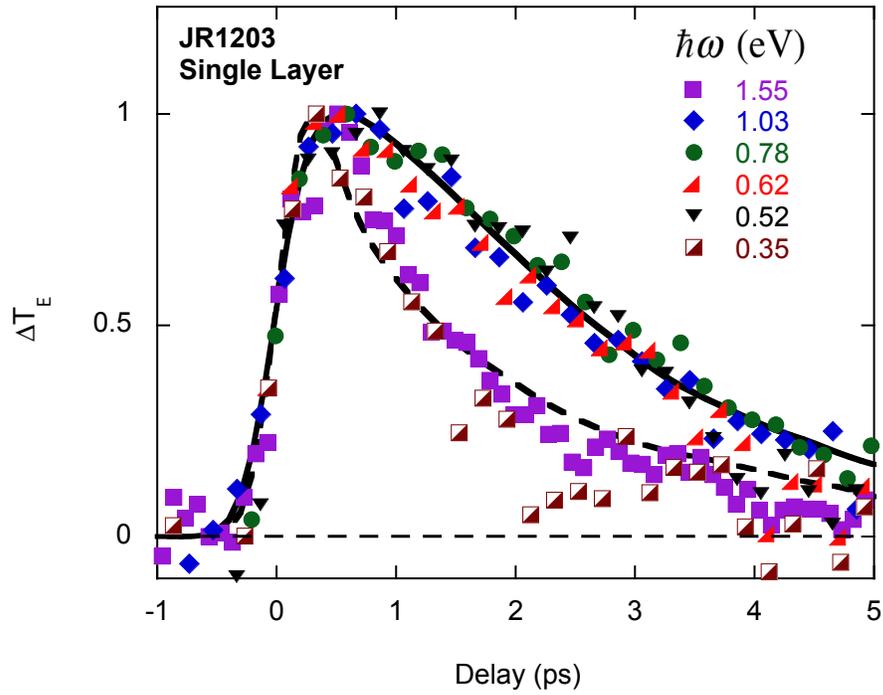

FIG. 3 Time-resolved THz transmission measurements on a single layer CVD graphene sample for a range of pump photon energies. Curves are normalized. Lines are model fits for absorbed pump intensities of 400 nJ/cm$^2$ (solid) and 40 nJ/cm$^2$ (dashed).

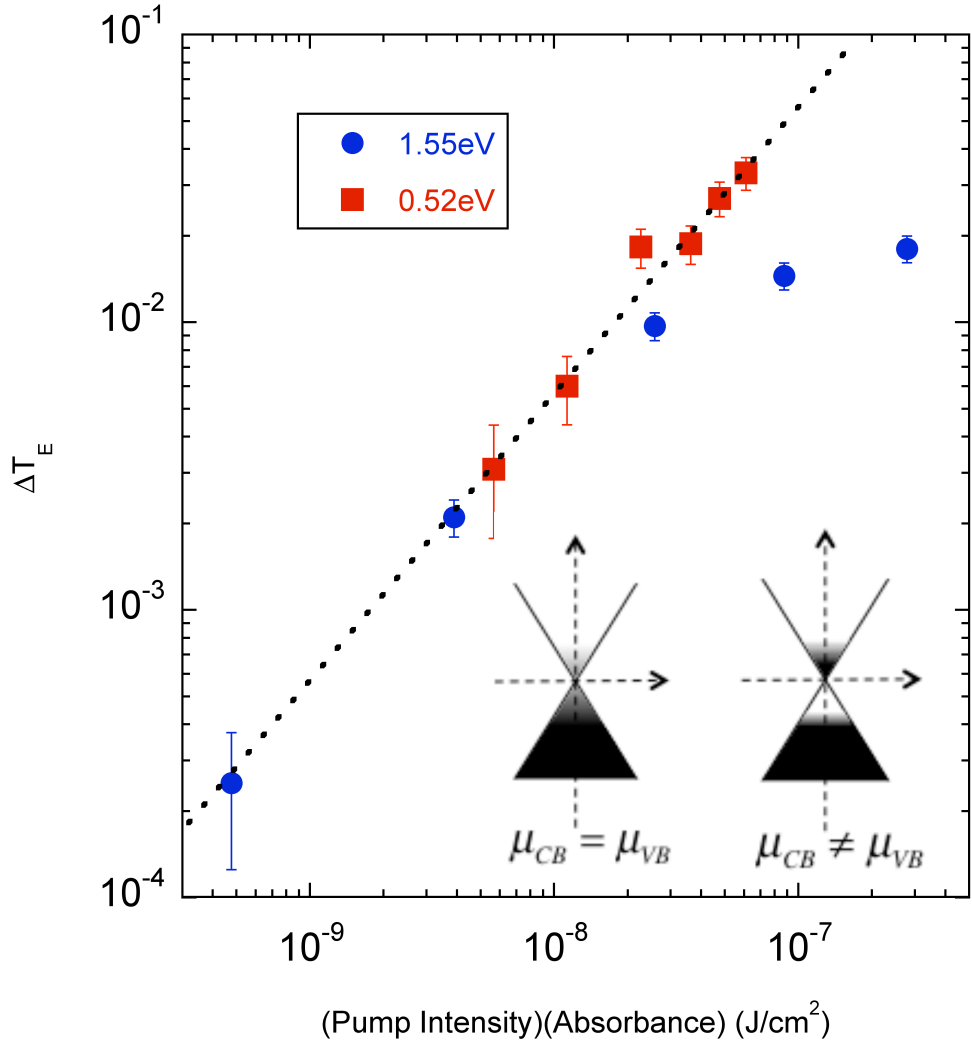

FIG. 4 Peak transmission change versus product of linear absorbance and pump energy density for pump photon energies 1.55eV ($\hbar\omega \gg 2\varepsilon_F$) and 0.52eV ($\hbar\omega \ll 2\varepsilon_F$). Dashed line shows linear dependence. Inset illustrates carrier distributions with (*i.*) a single carrier temperature and chemical potential and (*ii.*) distinct chemical potentials in the valance and conduction bands.